# A Simple Multi-Processor Computer Based on Subleq

*Oleg Mazonka and Alex Kolodin*
mazonka@gmail.com
alex.kolodin@gmail.com

Abstract

*Subleq (Subtract and Branch on result Less than or Equal to zero) is both an instruction set and a programming language for One Instruction Set Computer (OISC). We describe a hardware implementation of an array of 28 one-instruction Subleq processors on a low-cost FPGA board. Our test results demonstrate that computational power of our Subleq OISC multi-processor is comparable to that of CPU of a modern personal computer. Additionally, we provide implementation details of our complier from a C-style language to Subleq.*

## Contents

## *1. Introduction*

OISC (One Instruction Set Computer) is the ultimate RISC (Reduced Instruction Set Computer) with conventional CPU, where the number of instructions is reduced to one. Having only one available processor instruction eliminates the necessity in op-code and permits simpler computational elements, thus allowing more of them implemented in hardware with the same number of logical gates. Since our goal was to build a functional multi-processor system with a maximum possible number of processors on a single low-cost programmable chip, OISC was a natural choice, with the remaining step being selection of a suitable single-processor instruction set.

Currently known OISC can be roughly separated into three broad categories:

1. Transport Triggered Architecture Machines;
2. Bit Manipulating Machines;
3. Arithmetic Based Turing-Complete Machines.

*Transport Triggered Architecture* (TTA) is a design in which computation is a side effect of data transport. Usually some memory registers (triggering ports) within common address space, perform an assigned operation when the instruction references them. For example, in an OISC utilizing a single memory-to-memory copy instruction [1], this is done by triggering ports performing arithmetic and instruction pointer jumps when writing into them. Despite appealing simplicity, there are two major drawbacks associated with such OISC. First is that CPU has to have separate functional units controlling triggering ports. The second are difficulties with generalization of the design, since any two different hardware designs are likely to use two different assembly languages. Because of these disadvantages we ruled out this class of OISCs for our implementation.

*Bit Manipulating Machines* is the simplest class. Bit copying machine, called BitBitJump, copies one bit in memory and passes the execution unconditionally by the address specified by one of the operands of the instruction [2]. This process turns out to be capable of universal computation (i.e. being able to execute any algorithm and to interpret any other universal machine) because copying bits can conditionally modify the code ahead to be executed. Another machine, called Toga computer, inverts a bit and passes the execution conditionally depending on the result of inversion [3]. Yet another bit operating machine, similar to BitBitJump, copies several bits at the same time. The problem of computational universality is solved in this case by keeping predefined jump tables in the memory [4]. Despite simplicity of the bit operating machines, we ruled them out too, because they require more memory than is normally available on an inexpensive FPGA. To make a functional multiprocessor machine with bit manipulation operation, at least 1Mb of memory per processor is required. Therefore we decided that more complex processor with less memory is a better choice for our purposes.

*Arithmetic based Turing-complete Machines* use an arithmetic operation and a conditional jump. Unlike two previous classes which are universal computers, this class is universal *and* Turing-complete in its abstract representation. The instruction operates on integers which may also be addresses in memory. Currently there are several known OISCs of this class, based on different arithmetic operations [5]: addition – Addleq, decrement – DJN, increment – P1eq, and subtraction – Subleq (Subtract and Branch on result Less than or Equal to zero). The latter is the oldest, the most popular and, arguably, the most efficient [6][7]. A Subleq instruction consists of three operands: two for subtraction and one for conditional jump.

Attempts to build hardware around Subleq have been undertaken previously. For example, David A Roberts designed a Subleq CPU and wrote a software Subleq library [8]. His implementation is a single CPU with keyboard input, terminal, control ROM, and 16Mb RAM, and is much more complex than ours. There were a few other similar designs described on various Internet sites, e.g. [9]. However all of them were just proof-of-concept simulations without a practical implementation.

In the following sections we describe components of the system we built. Section 2 outlines Subleq abstract machine and its assembly notation. Section 3 describes our hardware implementation of the multiprocessor core. Section 4 briefly describes techniques used to convert high level programming language into Subleq instruction code. In Sections 5 and 6 we present comparative speed-test results for our device, followed by a discussion and summary. In the Appendix the code calculating factorials is presented in C and Subleq notations.

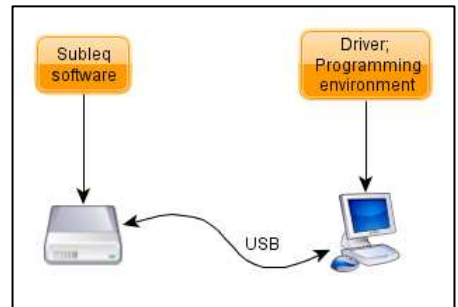

**Figure 1** FPGA board is connected to PC by USB cable

Figure 1 represents the connection of our device to a computer with USB cable.

## *2. Subleq Assembly Language*

Subleq abstract machine operates on an infinite array of memory, where each cell holds an integer number. This number can be an address of another memory cell. The numbering starts from zero. Program is defined as a sequence of instruction read from the memory with the first instruction at the address zero. A Subleq instruction has 3 operands:

        A B C

Execution of one instruction `A B C` subtracts the value in the memory cell at the address stored in `A` from the content of a memory cell at the address stored in `B` and then writes the result back into the cell with the address in `B`. If the value after subtraction in `B` is less or equal to zero, the execution jumps to the address specified in C; otherwise execution continues to the next instruction, i.e. the address of the memory cell following C.

Assembly notation helps to read and write code in Subleq. The following is the list of syntax conventions:

- label;
- question mark;
- reduced instruction;
- multi-instruction;
- literal and expression;
- data section;
- comment.

Label is a symbolic name of a particular address followed by a colon. In the following example

```
    A   B   C
A:2 B:1  0
C:B  B   0
```

each line represents one instruction with three operands. Here A, B, and C are not abstract names, but labels (addresses) of specific cells in the memory. For example, label A refers to the 4th cell, which is initialised with value 2. The first instruction subtracts value of the cell A from the value of the cell B, which value is 1, and stores the result in the cell B, which becomes -1. Since the result is less than zero, the next instruction to be executed is the third line, because value C is the address of the first operand of the instruction on the third line. That subtracts B from B making it zero, so the execution is passed to the address 0. If these three lines is the whole program, then the first operand of the first instruction has the address 0. In this case the execution is passed back to the first instruction which would make B -2. That process continues forever. The instructions being executed are only the first and the third lines, and the value of the cell B changes as 1, -1, 0, -2, 0, -2, 0, and so on.

Question mark is defined as the address of the next cell in memory:

```
A B ?
B B 0
```

is the same as

```
A B C
C:B B 0
```

Reduced instruction format is a convenient shortcut: two operands instead of three assume the third to be the address of next instruction, i.e. ?; and only one operand assumes the second to be the same as the first, so

```
A
```

is the same as

```
A A
```

and is the same as

```
    A A ?
```

If more than one instruction is placed on the same line, each instruction except the last must be followed by semicolon. The following code copies the value from A to B:

```
    Z; B; A Z; Z B
```

Integer numbers like `A:72` are used as constants in the code. Literals can be used instead of integers assuming their ASCII values. For example, `A:'H'` is the same as `A:72`; and `A:"Hi"` is the same as `A:'H' 'i'`. Addition, subtraction, parenthesis, and unary minus can be used in expression. The code

```
    Z Z ?+3
    A B C
    D E F
```

sets Z to zero and jumps to the third instruction D E F.

Since instructions can be reduced, the assembler must know when to generate full instruction with three operands. To avoid such generation, period at the beginning of the line is used. Thus program data can be placed on such line. The code

```
    A A ?+1
    . U:-1
    U A
```

sets A to 1. Without the period on the second line the code would be interpreted as

```
    A A ?+1
    U:-1 (-1) ?
    U A
```

Comments are delimited by hash symbol #: everything from # till the end of the line is a comment. Jump to a negative address halts the program. Usually (-1) as the third operand is used to stop the program, for example:

```
    # halt
    Z Z (-1)
```

The parentheses around (-1) are necessary to indicate that it is the third operand, so the instruction would not be interpreted as

```
    Z Z-1 ?
```

To make a Subleq program interactive (requesting data and responding to user while working), input and output operations can be defined as operations on a non-existing memory cell. The same (-1) address can be used for this. If the second operand is (-1) the value of the first operand is the output. If the first operand is (-1), the second operand gets the value from the input stream. Input and output operations are defined on byte basis in ASCII code. If the program tries to output a value greater than 255, the behaviour is undefined.

Below is a "Hello world" program adapted from Lawrence Woodman

`helloworld.sq` [10]. It is exceptionally terse, but is a good example of Subleq efficiency.

```
L:H (-1); U L; U ?+2; Z H (-1); Z Z L
 . U:-1 H:"hello, world\n" Z:0
```

A special variable called `z` is often used in Subleq as an intermediate temporary variable within a highly small scope. It is commonly assumed that this variable is initialised at zero and left at zero after every usage.

The program above consists of five instructions. First instruction prints the character pointed by its first operand (*the first pointer*) which is initialised to the beginning of the data string – the letter `'h'`. Second instruction increments that pointer – the first operand of the first instruction. Third instruction increments *the second pointer*, which is the second operand of the forth instruction. Forth instruction tests the value pointed by the second pointer and halts the program when the value is zero. It becomes zero when the pointer reaches the cell one after the end of the data string, which is `z:0`. The fifth instruction loops back to the beginning of the program, so the process continues until the halt condition is not satisfied.

## 3. Hardware design

### 3.1 Overview

We have used Altera Cyclone III EP3C16 FPGA as the basis for hardware implementation. The choice was based on relatively low price (~ US$30) of this FPGA IC chip and availability of the test hardware for it.

The test board we used has a DDR2 RAM IC fitted, but access to the RAM is limited by one process at a time. True parallel implementation requires separate internal memory blocks allocated for each processor hence the amount of available memory in the FPGA limits the number of processors. The EP3C16 FPGA has 56 of 16 bit memory blocks of 8 Kbits each. Our implementation of one 32 bit Subleq processor requires minimum 2 memory blocks, so only 28 processors can be fit into the FPGA. We could choose 16 bit implementation and have more processors (up to 56), but with only 1 Kbyte of memory allocated to each.

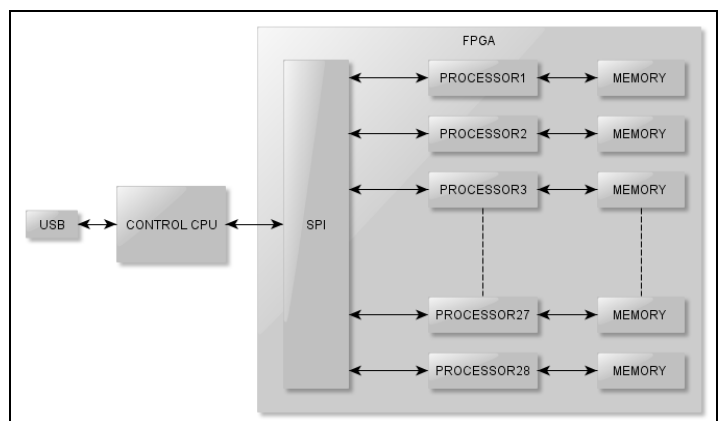

**Figure 2** Block-diagram of the board

FPGA is connected to USB bus with the help of external Cypress FX2 CPU that is configured as a bridge between USB and SPI (Serial Peripheral Interface) utilised to load FPGA with code and data. The interface bridge is transparent for the PC software.

Figure 2 is a communication block diagram of the board. The solution was coded in VHDL and compiled with Quartus II Web Edition software freely available from Altera website. Our code is scalable for up to 63 processors for possible use with larger FPGAs. The limit of 63 processors is due to our SPI bus addressing implementation and can be increased, if necessary.

All 28 processors run independently and synchronized by a single 150 MHz clock generated by one of the FPGA PLLs from a reference oscillator fitted on the PCB.

To increase the number of processors, additional boards with FPGAs could be easily connected via USB bus.

## 3.2 Interface description

Each processor has a serial interface to the allocated memory and the Status byte, accessible from a single address serial loading. The serial interface takes over memory's data and address buses when processing is stopped.

Address space inside the FPGA is organised as a table addressed by two numbers: processor index and memory address. Reading one byte from index 0 returns the number of processors inside the FPGA. For this design the returned value is 28. Indices from 1 to 28 are assigned to the processors, which have 2048 bytes (512 of 32 bit words) of memory available to each.

Writing to a processor memory is an operation which sequentially loads a buffer of 2048 bytes. Reading from the processor's memory is different: the first word (4 bytes) returned is the status of the processor and the rest is the memory content.

The Status byte – the first byte of the first word – can be in one of three states: `0xA1` – running, `0xA2` – stopped, or `0xA0` – stopped and not run since power on. Writing into the processor's memory automatically starts execution, thus eliminating the need in a separate command. Reading from a processor's memory stops that processor. An exception is reading the first byte of status, which does not stop the processor. Additionally, a processor can be stopped by Subleq halt operand (-1) as mentioned in Section 2. Other negative references, such as input or output as described above in the Subleq assembly language section, also stop the processor, because no IO operations are defined in this architecture.

## 3.3 Subleq processor

The state machine algorithm can be presented in pseudocode as:

```
IP = 0
while (IP >= 0)
{
        A = memory[IP]
        B = memory[IP+1]
        C = memory[IP+2]
        if (A < 0 or B < 0):
        {
                IP = -1
```

```
                    }
                    else:
                    {
                            memory[B] = memory[B] - memory[A]
                            if (memory[B] > 0)
                                    IP = IP + 3
                            else:
                                    IP = C
                    }
            }
```

where `IP` is an instruction pointer, `memory[]` is a value of a memory cell, and `A`, `B`, and `C` are integers.

The Subleq processor core is written with the help of RAM 2 Port Megafunction of Quartus II software we used to build dual port memory access. The implemented solution allows access to content by two distinct addresses (`memory[A]` and `memory[B]`) simultaneously, which results in saving on processing clock ticks. The disadvantage of this implementation is an additional latency of one clock tick to access the data and address buses comparing to a single port memory implementation. However the total of processing clock ticks per memory access for the dual port is still less than that required for a single port.

The core is based on a state machine, which starts automatically when memory of a processor is loaded. On any read or write operation, or encountering a negative operand the processing stops, but the first status byte can be read at any time without affecting the computation.

## *4. C Compiler for Subleq*

In this section we briefly describe some elements of the compiler we wrote, which compiles simplified C code into Subleq [11]. The compiler is used in one of our tests, so that direct comparison is possible between execution of a compiled native C code and a Subleq code, compiled from the same C source. The compiler is a high-level language interface to OISC – the only known to us such compiler at the time of writing.

### 4.1 Stack

The primary C programming language concepts are functions and the stack. Implementation of the stack in Subleq is achievable by using memory below the code. Using code self-modification one can place into and retrieve from the stack values. Function calls require return address to be placed into the stack. Consider the following C code:

```
void f()
{
        ...
}

void main()
{
        ...
```

```
        f();
        ...
    }
```

After the above is compiled to machine code, it must perform the following operations 1) the address of the instruction immediately after calling `f` has to be put on the stack; 2) a jump to the code of the function `f` must be made 3) at the end of the function `f`, the address from the stack needs to be extracted and 4) the execution should be transferred to the extracted address. According to C standard, the function `main` is a proper C function, i.e. it can be called from other functions including itself. Hence the program must have a separate entry point, which in the following code is called `sqmain`. The C code above compiles into:

```
            0 0 sqmain
    _f:
            ...
            #return
            ?+8; sp ?+4; ?+7; 0 ?+3; Z Z 0
    _main:
            ...
            #call f
            dec sp; ?+11; sp ?+7; ?+6; sp ?+2; 0
            ?+6; sp ?+2; ?+2 0 _f; . ?; inc sp
            ...
            #return
            ?+8; sp ?+4; ?+7; 0 ?+3; Z Z 0
    sqmain:
            #call main
            dec sp; ?+11; sp ?+7; ?+6; sp ?+2; 0
            ?+6; sp ?+2; ?+2 0 _main; . ?; inc sp
            0 0 (-1)
    . inc:-1 Z:0 dec:1 sp:-sp
```

The cell stack pointer `sp` is the last memory cell in the program. It is initialised with the negative value of its own address. Negative value is used here to speed up the code execution – working with subtraction operation may sometimes save a few steps if the data is recorded as negative of its actual value. The instruction `dec sp` subtracts 1 from `sp`, hence increasing its real value by 1. Below is an excerpt calling the function `f` in more readable form – relative references `?` are replaced with labels.

```
        dec sp
        A; sp A
        B; sp B
        A:0 B:0
        C; sp C
        D C:0 _f
        . D:?
        inc sp
```

The instruction on the forth line is to clear the cell in the stack, since some values can be left there from the previous use. However, to clear the top cell in the stack is not an single-step task because one has to clear the operands of the instruction itself and then initialise it with the value of `sp` pointer. Thus the execution command sequence is the following: allocate a new cell in the stack by increasing stack pointer (first line); clear the first operand of the instruction; initialise this operand with the new value of the stack pointer (second line); do the same with the second operand of the instruction – clear and initialise (third line); and then execute the instruction, which will clear the allocated cell in the stack (forth line).

The next two instructions clear and initialise cell C similarly. The instruction `D C:0 _f` copies the address of the instruction `inc sp` to the stack and jumps to `_f`. This works, because `D` holds the value of the next memory cell (remember `?`) and `C` points to the now cleared top cell on the stack. A negative value written to the stack then forces a jump to the label `_f`.

Once inside the function `f`, stack pointer can be modified, but we assume that functions restore it before they exit. So the return code has to jump to the address extracted from the stack:

```
A; sp A
B; A:0 B
Z Z B:0
```

Here the value of stack pointer `sp` is written to `A`, and the instruction `A:0 B` copies the stored address to `B`. The address stored negatively so the positive value is being restored.

The stack does a little bit more than just storing return address. That will be explored later in subsections 4.3 and 4.4.

## 4.2 Expressions

C language operations consist of statements, which in turn consist of keyword statements and expressions. The syntax of keyword statements and expressions are best represented by Backus-Naur Forms (BNF) – the standard way of representing context-free grammars. A classic example is the grammar of arithmetic expressions:

```
expression:=
        term
        expression + term
        expression - term
term:=
        primary
        term * primary
        term / primary
primary:=
        identifier
        constant
        ( expression )
```

This mutually recursive definitions can be used by a program called parser to build a tree representation for any grammatically valid expression. Once such tree is build, the compiler job is to organise the sequence of instructions so that the result of any sub-tree is passed up the tree. For example, a tree of the expression:

```
a + ( b - c )
```

consists of a node '+', variable `a` and a sub-tree, which consists of a node '-', and variables `b` and `c`. To make calculation, the compiler must use a temporary variable to store the result of the sub-tree, which has to be used later in addition; and potentially to be used further up if this expression is a part of a larger expression. In this particular example we need only one temporary, but generally many temporaries are required. The expression is compiled into the following code:

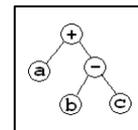

```
t; b Z; Z t; Z
c t
a Z; Z t; Z
```

The first line copies value `b` into temporary `t`. The second line subtracts value `c` from the temporary. At this point the compiler is finished with sub-tree. Its result is the generated code and the temporary variable `t` holding the value of the calculated sub-tree. Now compiler generates code for the addition. Its arguments now are variable `a` and temporary `t`. The third line adds `a` to `t`. Now `t` holds the result of the whole expression. If this expression is a part of a larger expression, `t` is passed up the tree as an argument to the upper level node of the tree. If not, then `t` value is discarded because the evaluation is finished.

More advanced grammar may involve assignment, dereferencing, unary operations and many others. But each grammar construction can be represented by a corresponding sub-tree, and processed later by the compiler to produce code. For example, a subtraction from a dereferenced value represented in C as:

```
*k -= a
```

has to be translated into

```
t; k Z; Z t; Z
a t:0
```

Here a temporary variable has to be used inside the code for dereferencing. The sequence of instructions is: clear `t`; copy value `k` into `t`; subtract `a` from the memory `k` points to.

Here a few elements of the grammar processing have been touched. C grammar takes several pages just to list the BNF. However larger and more complex grammars are resolved by the compiler is similar manner.

## 4.3 Function calls

In the subsection 4.1 above it was shown how to push into and pop from the stack. When a function takes arguments, they have to be pushed into the stack together with the return address. The stack must be restored upon function return. Consider a function taking two arguments:

```
int f(int a, int b);
        ...
        f(a,b);
```

The call to a function `f` must be translated into something like this

```
# 1 push b
# 2 push a
# 3 push return_address
# 4 goto f
# return_address:
# 5 sp -= 3
```

In C arguments can be expressions and the call to a function can be a part of another expression - sub-expression, i.e. the compiler must properly handle more complicated cases like the following

```
int f(int a, int b)
{
  ...
  return f;
}
  ...
  int k;
  k=f;
  k(f(1,2),3);    // call via variable - indirect call
  k = f(1,2)(3,4); // call by return value
```

Here for simplicity C function type `int(*)(int,int)` is represented as `int`. Subleq supports only one variable type. Therefore, more elaborate typing system does not introduce extra functionality in the language.

Arguments pushed into the stack can properly be calculated as sub-expressions (subtree). In this sense for the actual function call it is irrelevant either program variables or temporary are pushed into the stack.

```
# 1 push B
  # clearing the next cell in the stack [remember that sp is negative]
  # the line below is same as in C syntax: *(++sp)=0;
  dec sp; t1; sp t1; t2; sp t2; t1:0 t2:0
  # same as in C syntax: *sp+=B;
  t3; sp t3; b Z; Z t3:0; Z

# 2 push A
  # the same with A
  dec sp; t4; sp t4; t5; sp t5; t4:0 t5:0
  t6; sp t6; a Z; Z t6:0; Z

# 3 push return_address
  dec sp; t7; sp t7; t8; sp t8; t7:0 t8:0
  t9; sp t9; t10 t9:0 goto_address
  . t10: return_address

# 4 goto f
  goto_address: Z Z f

# 5 sp -= 3
  return_address: const(-3) sp
```

Notation `const(-3) sp` is a short for

```
  unique_name sp
  ...
  unique_name:-3
```

The code above handles neither return value nor indirect calls yet. Return value can be stored in a special variable (register). If the program uses the return value in a sub-expression, then it must copy the value into a temporary immediately upon return. Indirect calls can be achieved by dereferencing a temporary holding the address of the function. It is straightforward, but more complex code.

Stack pointer can be modified inside a function when the function requests stack (local) variables. For accessing local variables usually base pointer `bp` is used. It is initialised on function entrance; is used as a base reference for local variables – each local variable has an associated offset from base pointer; and is used to restore stack pointer at the end of the function. Functions can call other functions, which means that each function must save upon entry and restore upon exit base pointer. So the function body has to be wrapped with the following commands:

~ 12 ~

```
1. # push bp
2. # sp -> bp
3. # sp -= stack_size

   # ... function body

5. # bp -> sp
6. # pop bp
7. # return
```

Or in Subleq code.

```
dec sp; ?+11; sp ?+7; ?+6; sp ?+2; 0
?+6; sp ?+2; bp 0
bp; sp bp
stack_size sp

# ... function body

sp; bp sp
?+8; sp ?+4; bp; 0 bp; inc sp
?+8; sp ?+4; ?+7; 0 ?+3; Z Z 0
```

`stack_size` is a constant, which is calculated for every function during parsing. It turns out that it is not enough to save `bp`. A function call can happen inside an expression. In such case all temporaries of the expression have to be saved. A new function will be using the same temporary memory cells for its own needs. For the expression `f()+g()` the results of the calls may be stored in variables `t1` and `t2`. If function `g` changes `t1` where the result of function `f` is stored, a problem would appear.

A solution is to make every function push all temporaries it is using into the stack and to restore them upon exit. Consider the following function:

```
int g()
{
  return k+1;
}
```

It translates into:

```
_g:
  # save bp
  dec sp; ?+11; sp ?+7; ?+6; sp ?+2; 0
  ?+6; sp ?+2; bp 0
  bp; sp bp

  # push t1
  dec sp; ?+11; sp ?+7; ?+6; sp ?+2; 0
  ?+6; sp ?+2; t1 0
  # push t2
  dec sp; ?+11; sp ?+7; ?+6; sp ?+2; 0
  ?+6; sp ?+2; t2 0

  # calculate addition
  t1; t2
  _k t1
  dec t1
  t1 t2
  # set the return value [negative]
  ax; t2 ax

  # pop t2
  ?+8; sp ?+4; t2; 0 t2; inc sp
  # pop t1
  ?+8; sp ?+4; t1; 0 t1; inc sp
```

```
               # restore bp
               sp; bp sp
               ?+8; sp ?+4; bp; 0 bp; inc sp
               # exit
               ?+8; sp ?+4; ?+7; 0 ?+3; Z Z 0
```

If somewhere inside the code there are calls to other functions, the temporaries $t1$ and $t2$ hold their calculated values because other functions save and restore them when executed.

Since all used temporaries in the function are pushed into the stack, it pays off to reduce the number of used temporaries. It is possible to do this just by releasing any used temporary into a pool of used temporaries. Then later when a new temporary is requested, the pool is first checked and a new temporary is allocated only when the pool is empty.

The expression

```
               1+k[1]
```

compiles into

```
               t1; t2; _k t1; dec t1; t1 t2
               t3; t4; ?+11; t2 Z; Z ?+4; Z; 0 t3; t3 t4;
               t5; t6; dec t5; t4 t5; t5 t6
               # result in t6
```

When pool of temporaries is introduced the number of temporaries is halved:

```
               t1; t2; _k t1; dec t1; t1 t2
               t1; t3; ?+11; t2 Z; Z ?+4; Z; 0 t1; t1 t3
               t1; t2; dec t1; t3 t1; t1 t2
               # result in t2
```

which dramatically reduces the code removing corresponding push and pop operations.

### 4.4 Stack variables

Once `bp` is placed on the stack and `sp` is decremented to allocate memory, all local variables become available. They can be accessed only indirectly because the compiler does not know their addresses. For example, the function `f` in

```
               int f(int x, int y)
               {
                 int a, b=3, c[3], d=5;
                 ...
               }
                 f(7,9);
```

has 4 local variables with the stack size equal to 6. When this function is entered the stack has the following values:

```
... y[9] x[7] [return_address] [saved_bp] a[?] b[3] c₀[?] c₁[?] c₂[?] d[5] ...
                                    ^                                    ^
                                   (bp)                                 (sp)
```

The compiler knows about the offset of each variable from `bp`.

| Variable | Offset |
|----------|--------|
| y        | -3     |
| x        | -2     |
| a        | 1      |
| b        | 2      |
| c        | 3      |
| d        | 6      |

Hence, in the code any reference to a local variable, not pointing to an array, can be replaced with `*(bp+offset)`. The array `c` has to be replaced with `(bp+offset)` because the name of array is the address of its first element. The name does not refer to a variable, but the referencing with `[]` does. In C

```
c[i]
```

is the same as

```
*(c+i)
```

which can be interpreted in our example as

```
*((bp+3)+i)
```

## 4.5 Multiplication

The only trivial multiplication in Subleq is multiplication by 2, `t=a+a`:

```
t; a Z; a Z; Z t; Z
```

To multiply 2 numbers one can use formula

```
A*B = (2A)*(B/2) + A*(B%2)
```

This is a simple recursive formula, but it requires integer and modular division. Division can be implemented as the following algorithm. Given two numbers A and B, B is increased by 2 until the next increase gives B greater then A. At the same time as increasing B, we increase another variable I by 2, which has been initialized to 1. When B becomes greater then A, I holds the part of the result of division – the rest is to be calculated further using A-B and original B. This can be done recursively accumulating all I's. At the last step when A<B, A is the modulus. This algorithm can be implemented as a short recursive function in C. Upon the exit this function returns the integer division as the result and division modulus in the argument j.

```
int divMod(int a, int b, int * j)
{
   if( a < b ) { *j=a; return 0; }
   int b1=b, i=1, bp, ip;
 next:
   bp = b1; ip = i;
   b1 *= 2; i *= 2;
```

```
    if( b1 > a )
        return ip+divMod(a-bp,b,j);

    goto next;
}
```

This function is not optimal. More efficient function can be achieved by replacing recursion with another external loop. Multiplication, integer and modular division operations requiring quite elaborate calculations can be implemented as library functions. That is, each multiplication `a*b` can be replaced with a call `_mul(a,b)`, and later the compiler may add (if necessary) the implementation of the function.

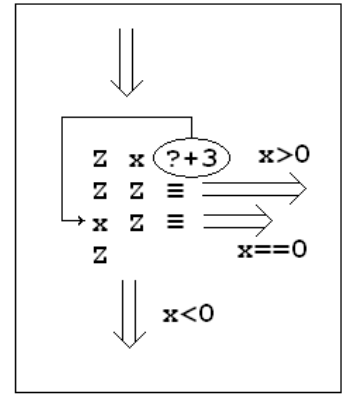

**Figure 3** Diagram representing conditional jumps.

## 4.6 Conditional jump

In C, Boolean expressions which evaluate to zero are **false** and non-zero are **true**. In Subleq this leads to longer code when handling Boolean expressions because every Boolean expression evaluates on basis of equality or non-equality to zero.

A better approach is to treat less or equal to zero as **false** and positive value as **true**. Then if-expression `if(expr){<body>}` will be just one instruction

```
Z t next
<body>
next: ...
```

where `t` is the result of the expression `expr`. However to remain fully compatible with C (for example, `if(x+1){...}` – an implicit conversion to Boolean) all cases where integer expression is used as Boolean have to be detected. Fortunately there are only a few such cases:

- `if(expr)`
- `while(expr)`
- `for(...,expr,...)`
- `! expr`
- `expr1 && expr2`
- `expr1 || expr2`

The job can be done inside the parser, so the compiler would not have to care about Boolean or integer expression, and it can produce much simpler code.

In cases when a Boolean variable is used in expressions as integer, like in:

- passing an argument **f(a>0)**
- returning from a function **return(a>0);**
- assignment **x=(a>0);**
- other arithmetic expression **x=1+5*(a>0);**

the variable must be converted to C-style, i.e. negative result zeroed. This can be done as simple as

```
    x Z ?+3; x; Z
```

A terse check for a value being less than, equal to, or greater than zero is:

```
    Z x ?+3; Z Z G; x Z E; Z; L:
```

where `L`, `E`, and `G` are addresses to pass the execution in cases when `x` is less than, equal to, or greater than zero respectively. Figure 3 shows the schema of the execution. Note, that `x` does not change and `z` is zero on any exit!

## 5. Results

Figure 4 shows our FPGA board powered via USB cable. Sized about 5 x 7 centimetres the board implements 28 Subleq processors with allocated memory of 2 Kb per processor and running at clock frequency 150 MHz.

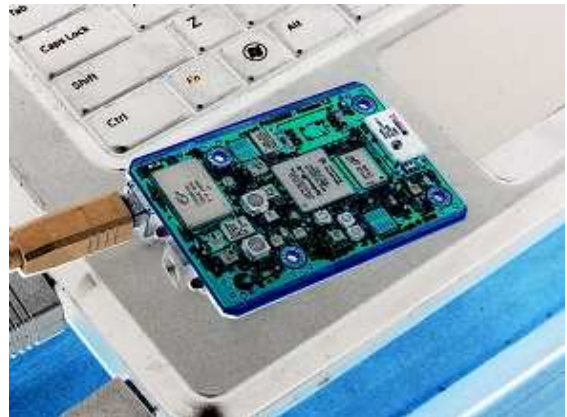

**Figure 4** FPGA board, 28 Subleq processors with allocated 2 Kb per processor

To test the efficiency of the board we chose two mathematical problems. The first calculates the size of a function residue of an arithmetic group. The second calculates modular double factorials.

### 5.1 Test #1

In the first test we selected a problem of finding order of function residue of the following process:

$$x_{i+1} = 2x_i \bmod M$$
$$y_{i+1} = 2(x_i + y_i) \bmod M$$

where *x* and *y* are integers initialised to 1, mod is a modulo operation, *M* is some value. Starting from the point ($x_0=1, y_0=1$) the equations generate a sequence of pairs. We chose this problem because its solution is difficult, with answers often much greater than *M* (but less than $M^2$). Number *M* was selected such, that the calculations could be completed in few minutes. When this sequence is sufficiently long, a new pair of generated numbers will eventually be the same as a pair previously generated in the sequence. The task is to find how many steps has to be completed before first occurrence of the result with the same value. In our test the selected value of *M* was *M*=5039 and the number of iterations was calculated as 12693241.

A C program to solve this problem can be written without use of multiplication or division:

```
int x=1, y=1, m=5039;
int c=0, ctr=1, t;
int x0=0, y0=0;

int printf();
int main()
{
    while(1)
    {
        y += x; y += y; x += x;
        while( x>=m ) x-=m;
        while( y>=m ) y-=m;

        if( x==x0 && y==y0 ) break;

        if( ++c==ctr )
        {
            x0=x; y0=y;
            c=0; ctr+=ctr;
        }
    }
    printf("point: %d %d  loop: %d of %d\n",x0,y0,c+1,ctr);
}
```

This program has been tested in the following cases:

1. Compiled with our Subleq compiler and run on one of the processors on FPGA board;
2. Compiled with our Subleq compiler and emulated on PC#1 (Intel Q9650 at 3GHz)
3. Compiled with Microsoft C/C++ compiler (v16) with full optimisation and run on PC#1.
4. Same as 2, but run on PC#2 (Pentium 4 at 1.7GHz)
5. Same as 3 run on PC#2

The table below shows execution time in seconds for each test.

| 1 | Subleq on 1 processor FPGA | 94.0 |
|---|---|---|
| 2 | Subleq on PC#1 | 46.0 |
| 3 | C on PC#1 | 0.37 |
| 4 | Subleq on PC#2 | 216 |
| 5 | C on PC#2 | 0.54 |

From these results we conclude that the speed of a single processor on FPGA is of the same order of magnitude as the speed of CPU of ordinary PC when emulating Subleq instructions. Native code on PC runs about hundred times faster.

## 5.2 Test #2

The second test was a calculation of modular double factorials, namely

$$(N!)! \bmod M = \prod_{n=1}^{N} \prod_{i=1}^{n} i \bmod M$$

In this test case we were able to use full power of our multi-processor Subleq system because multiplication in the above equation could be calculated in parallel across all 28 processors. For *N*=5029 and *M*=5039 the result is 95 and these numbers were used in the test. Number *M* was same as in the Test#1 and number *N* was selected to give the result (95) in ASCII printable range. The calculations were run in the following configurations:

1. Hand written Subleq code run of FPGA board [Appendix 7.3]
2. Subleq code emulated on PC (same as PC#1 in the first test)
3. Equivalent C code compiled with the same C compiler and run on PC [Appendix 7.1]
4. Same C code compiled with Subleq compiler and emulated on PC
5. Equivalent C code without multiplication operation compiled with C compiler and run on PC [Appendix 7.2]
6. Same C code as in 5 compiled with Subeq compiler and emulated on PC

The code we used not 100% efficient, since the solution to the problem needs ~$O(N\log N)$ operations if utilising modular exponentiation, rather than ~$O(N^2)$ as presented in the Apendix. However this is not important when evaluating relative performance.

The results are presented in the table below. The values are execution time in seconds.

| 1 | Subleq on FPGA, parallel on 28 processors | 62.0 |
| 2 | Subleq on PC (emulation) | 865 |
| 3 | C with multiplication, executable run on PC | 0.15 |
| 4 | C with multiplication, Subleq emulated on PC | 12060 |
| 5 | C without multiplication, executable run on PC | 7.8 |
| 6 | C without multiplication, Subleq emulated on PC | 9795 |

The 28 FPGA processors easily outperform the emulation of the same Subleq code on PC. C code without multiplication compiled into Subleq and emulated runs faster than C code with multiplication, because the compiler's library multiplication function is not as efficient as the multiplication function written in this example.

## *6. Conclusion*

Using inexpensive Cyclone III FPGA we have successfully built an OISC multi-processor device with processors running in parallel. Each processor has its own memory limited to 2 Kb. Due to this limitation we were unable to build a multi-processor board with even simpler individual processor instruction set, such as e.g. bit copying [2], because in that case, to run practically useful computational tasks the minimum required memory is ~ 1 Mb of memory per processor. The limited memory available in our device also did not permit us to run more advanced programs, such as emulators of another processor or use more complex computational algorithms, because all the computational code has to fit inside the memory allocated for each processor.

The size of memory available to each processor can be increased by choosing larger and faster albeit more expensive FPGA such as Stratix V. Then a faster processing clock and larger number of CPUs could be implemented as well. The VHDL code of the CPU state machine could also be optimised improving computational speed. Given sufficient memory, it would be possible to emulate any other processor architecture, and use algorithms written for other CPU's or run an operating system. Apart from the memory constrain, another downside of this minimalist approach was reduced speed. Our board uses rather slow CPU clock speed of 150 MHz. As mentioned above, more expensive FPGA can run at much faster clock speeds.

On the other hand, the simplicity of our design allows for it to be implemented as a standalone miniature-scale multi-processor computer, thus reducing both physical size and energy consumption. With proper hardware, it might also be possible to power such devices with low power solar batteries similar to those used in cheap calculators. Our implementation is scalable – it is easy to increase the number of processors by connecting additional boards without significant load on host's power supply. A host PC does not have to be fast to load the code and read back the results. Since our implementation is FPGA based, it is possible to create other types of runtime re-loadable CPUs, customised for specific tasks by reprogramming FPGA.

In conclusion, we have demonstrated feasibility of OISC concept and applied it to building a functional prototype of OISC multi-processor system. Our results demonstrate that with proper hardware and software implementation, a substantial computational power can be achieved already in a very simple OISC multi-processor design.

# 7. Appendix

This section presents pieces of code calculating modular double factorial.

## 7.1 C with multiplication

The following C program calculates modular double factorial using built-in multiplication and division operations.

```
1       int printf();
2       int main()
3       {
4               int a=5029;
5               int b=1;
6               int m=5039;
7               int x=1;
8               int i,j;
9
10              for( i=a; i>b; i-- )
11              for( j=1; j<=i; j++ )
12              x = (j*x)%m;
13
14              printf("%d",x);
15      }
```

Lines 10-12 is a double loop multiplying numbers from `b` to `a` modulo `m`.

## 7.2 C without multiplication

This C program does the same calculation as the program above in 7.1, but without built-in multiplication and division operations. Multiplication and division functions are written explicitly.

```
1       int DivMod(int a, int b, int *m)
2       {
3               int b1, i1, bp, ip;
4               int z = 0;
5
6       start:
7               if( a<b ){ *m=a; return z; }
8
9               b1=b; i1=1;
10
11      next:
12              bp = b1; ip = i1;
13              b1 += b1; i1 += i1;
14
15              if( b1 > a )
16              {
17                      a = a-bp;
18                      z += ip;
19                      goto start;
20              }
21
22              if( b1 < 0 ) return z;
23
24              goto next;
25      }
26

27      int Mult(int a, int b)
28      {
29              int dmm, r=0;
30
31              while(1)
32              {
33                      if( !a ) return r;
34                      a=DivMod(a,2,&dmm);
35                      if( dmm ) r += b;
36                      b += b;
37              }
38      }
39
40      int printf();
41
42      int a=5029, b=1, m=5039;
43      int k=0, x=1, t;
44
45      int main()
46      {
47      start:  k=a;
48      loop:   t=Mult(k,x);
49              DivMod(t,m,&x);
50
51              if( --k ) goto loop;
52              if( --a > b ) goto start;
53
54              printf("%d",x);
55      }
```

Lines 1-25 implement the division algorithm described in 4.5, but optimised by removing recursive call. The multiplication (lines 27-38) is a straightforward implementation of the formula shown in 4.5. C loops are replaced with goto statements to make process flow similar to Subleq implementation in the next subsection 7.3.

## 7.3 Subleq code

Subleq code calculating modular double factorials has been written manually, because the compiled Subleq from C did not fit into the memory. The code below has 83 instructions, which can fit even into 1 Kb with 32-bit word.

```
1       0 0 Start
2
3       . A:5029 B:1 MOD:5039
4       . Z:0 K:0 X:1
5
6       Start:
7       A Z; Z K; Z
8
9       Loop:
10      mu_a; K mu_a
11      mu_b; X mu_b
12
13
14      Mult:
15      mu_r
16
17      mu_begin:
18      t2; mu_a t2 mu_return:N2
19
20      dm_a; mu_a dm_a
21      dm_b; C2 dm_b
22      dm_return; N3L dm_return
23      t2 t2 DivMod
24
25      N3:
26      dm_m t2 ?+3
27
28      mu_b mu_r
29
30      mu_a; dm_z mu_a
31      mu_b Z; Z mu_b; Z Z mu_begin
32
33      . mu_a:0 mu_b:0 mu_r:0
34
35      #Mult
36
37
38      N2:
39      dm_a; mu_r Z; Z dm_a; Z
40      dm_b; MOD dm_b
41
42      dm_return; N1L dm_return
43      Z Z DivMod
44
45      N1:
46      X; dm_m X

47
48      C1 K ?+3
49      Z Z Loop
50
51      C1 A
52      B A END
53      B Z; Z A; Z
54      K K Start
55
56      END:
57      X̶ ̶(̶-̶1̶)̶
58      Z Z (-1)
59
60      DivMod:
61
62      dm_z
63      dm_m
64
65      dm_start:
66      t1; dm_b t1
67      dm_a t1 ?+6
68      dm_a dm_m; Z Z dm_return:0
69
70      dm_b1; dm_b Z; Z dm_b1; Z
71      dm_i1; C1 dm_i1
72
73      dm_next:
74      dm_bp; dm_b1 dm_bp
75      dm_ip; dm_i1 dm_ip
76
77      dm_b1 Z; Z dm_b1; Z
78      dm_i1 Z; Z dm_i1; Z
79      t1; dm_b1 t1
80      dm_a t1 dm_next
81
82      dm_bp dm_a
83      dm_ip Z; Z dm_z; Z Z dm_start
84
85      . dm_a:0 dm_b:0 dm_z:0
86      . dm_m:0 dm_b1:0 dm_ip:0
87      . dm_i1:0 dm_bp:0 t1:0
88
89      #divMod
90
91      . N1L:-N1 N3L:-N3 t2:0
92      . C1:1 C2:2 0
```

Lines 3 and 4 define variables similar to how variables are defined in the C example above. A defines the number for which double factorial is to be calculated; B is a starting number – in our case it is 1, but in general it can be any number. When the task is distributed among parallel processors the range B to A is broken into smaller ranges and submitted to the processors independently. Upon completion the results collected and processed further. MOD is modulus of the algorithm. Z is Subleq zero register. K is an intermediate value running from A to 1. And X is the accumulated result.

Line 7 initialises K. Lines 10 and 11 prepare formal arguments for the multiplication algorithm written between lines 14 and 35. This code of the multiplication algorithm is almost one to one equivalent to the function `Mult` written in the previous sub-section. The only complex is that `DivMod` function organised here is a function so its code is being reused from the lines 23 and 43. To make this possible one needs initialise formal arguments for the function as well as the return address. The return address is copied via indirect labels `N1L` and `N3L`.

Lines 39 and 40 take the result from multiplication and initialise arguments for division. Lines 42 and 43 initialise return address and call `DivMod`. Line 46 extracts the result into `x`.

Lines 48 and 49 decrement `K` and check if it is less than 1. If not, the whole iteration is repeated with `K` smaller by 1. If `K` reached zero, proceed.

Lines 51-54 decrement `A` and check if `A` has reached `B`. If yes, we go to label `END`. If not, we go to line 7 and repeat the whole process again but now with `A` reduced by, hence `K` starting from new value of `A`.

Line 57 crossed over is to print the result. This instruction is handy when emulating Subleq. But in calculating on the FPGA board this instruction does not exists because the board does not have concept of input-output operations. The next line 58 is a valid Subleq halt command.

Lines 60-89 are corresponding Subleq code for the function `DivMod` presented in C in the sub-section above.

Finally, lines 91 and 92 define return addresses for calls to `DivMod`, a temporary `t2`, and two constants 1 and 2. The later is required for division in the multiplication formula from 4.5.